\documentclass[%
 aip,
 amsmath,amssymb,
 reprint,%
]{revtex4-2}

\usepackage{graphicx}
\graphicspath{{figures/}}
\usepackage{dcolumn}
\usepackage{bm}

\usepackage[utf8]{inputenc}
\usepackage[T1]{fontenc}
\usepackage{siunitx}

\DeclareMathOperator{\sgn}{sgn}
\DeclareMathOperator{\re}{Re}
\DeclareMathOperator{\im}{Im}

\DeclareMathOperator{\erfi}{erfi}

\makeatletter
\def\@email#1#2{%
 \endgroup
 \patchcmd{\titleblock@produce}
  {\frontmatter@RRAPformat}
  {\frontmatter@RRAPformat{\produce@RRAP{*#1\href{mailto:#2}{#2}}}\frontmatter@RRAPformat}
  {}{}
}%
\makeatother
\begin{document}

\preprint{AIP/123-QED}


\title{Computing the generalized plasma dispersion function for non-Maxwellian plasmas, with applications to Thomson scattering}

\author{C. R. Skolar}
\affiliation{Center for Solar-Terrestrial Research, New Jersey Institute of Technology, Newark, NJ 07102, USA}
\email[]{chirag.skolar@njit.edu}

\author{W. J. Longley}
\email[]{william.longley@njit.edu}
\affiliation{Center for Solar-Terrestrial Research, New Jersey Institute of Technology, Newark, NJ 07102, USA}

\author{L. V. Goodwin}
\affiliation{Center for Solar-Terrestrial Research, New Jersey Institute of Technology, Newark, NJ 07102, USA}
\email[]{lindsay.v.goodwin@njit.edu}

\date{\today}

\begin{abstract}
Kinetic plasma studies often require computing integrals of the velocity distribution over a complex-valued pole. The standard method is to solve the integral in the complex plane using the Plemelj theorem, resulting in the standard plasma dispersion function for Maxwellian plasmas. For non-Maxwellian plasmas, the Plemelj theorem does not generalize to an analytic form, and computational methods must be used. In this paper, a new computational method is developed to accurately integrate a non-Maxwellian velocity distribution over an arbitrary set of complex valued poles. This method works by keeping the integration contour on the real line, and applying a trapezoid rule-like integration scheme over all discretized intervals. In intervals containing a pole, the velocity distribution is linearly interpolated, and the analytic result for the integral over a linear function is used. The integration scheme is validated by comparing its results to the analytic plasma dispersion function for Maxwellian distributions. We then show the utility of this method by computing the Thomson scattering spectra for several non-Maxwellian distributions: the kappa, super Gaussian, and toroidal distributions. Thomson scattering is a valuable plasma diagnostic tool for both laboratory and space plasmas, but the technique relies on fitting measured wave spectra to a forward model, which typically assumes Maxwellian plasmas. Therefore, this integration method can expand the capabilities of Thomson scatter diagnostics to regimes where the plasma is non-Maxwellian, including high energy density plasmas, frictionally heated plasmas in the ionosphere, and plasmas with a substantial suprathermal electron tail.
\end{abstract}

\maketitle

\section{Introduction}
The plasma dispersion function\cite{fried1961} is typically defined for a Maxwellian distribution as 
\begin{equation}
    Z_M(z) = \frac{1}{\sqrt{\pi}} 
    \int_{-\infty}^\infty \frac{\exp(-v^2)}{v-z} dv
    \label{eq:plasma-dispersion-function}
\end{equation}
for $\im(z)>0$ and is analytically continued for $\im(z) \leq 0$.
However, for instances where the plasma is described by a non-Maxwellian distribution function, the plasma dispersion function is generalized to
\begin{equation}
    Z(z) = \int_{-\infty}^\infty \frac{f(v)}{v-z} dv,
    \label{eq:gen-plasma-dispersion-function-one-pole}
\end{equation}
where $f(v)$ is an arbitrary distribution function. 
The denominator of both Eqs.~\ref{eq:plasma-dispersion-function} and \ref{eq:gen-plasma-dispersion-function-one-pole} has a first order pole at $z$, leading to difficulties in evaluating the integral. A full computation of the plasma's dielectric function or Thomson scattering spectra also involves integrating the distribution function over other types of poles. Therefore, we define for this paper the generalized plasma dispersion function for an arbitrary set of poles:
\begin{equation}
    Z(z_1, z_2, \ldots z_i, \ldots, z_N) = \int_{-\infty}^\infty \frac{f (v)}{\prod_{i=0}^N (v - z_i)^{r_i}}dv,
    \label{eq:gen-plasma-dispersion-function-sum-poles}
\end{equation}
where the denominator is a product of $N$ poles, $z_i$, each with order $r_i$. For each pole, $z_i$ is arbitrary, but must have a nonzero imaginary part. This requirement is automatically satisfied when the kinetic equations are Laplace transformed in time, including the case where the imaginary part of the pole limits to 0 for collisionless damping. Furthermore, the list of poles $z_i$ can include complex conjugates, as these will be treated as different poles.  In this paper, we develop a novel method for numerically integrating Eq. \ref{eq:gen-plasma-dispersion-function-sum-poles}. 

This paper is structured so that readers with a wide range of backgrounds and use cases can immediately follow the development of the numerical integration scheme for Eq. \ref{eq:gen-plasma-dispersion-function-sum-poles} in Sec. \ref{sec:method}.  The motivation for this work is to calculate Thomson scattering spectra for non-Maxwellian plasmas. As the numerical scheme developed in Sec. \ref{sec:method} applies broadly to any kinetic theory calculations, the motivation and application to Thomson scattering are not discussed until Sec. \ref{sec:thomson_app}. Sec. \ref{sec:thomson_results} shows the calculation of Thomson scattering spectra for several non-Maxwellian distribution functions, including kappa, super Gaussian, and toroidal distributions. Finally, Sec. \ref{sec:conclusions} provides a summmary and the conclusions of this work. 

Compared to other modern techniques, the novel (yet simple) method presented here provides faster and more accurate calculations for any set of poles.

\section{Previous Methods}

The plasma dispersion function is a standard term for describing wave properties in a plasma, and there are many decades of literature on its derivation and computation, including standard textbooks.\cite{nicholson, bellan, froula} We briefly review the standard solution which invokes the Plemelj theorem before developing our new numerical method below. The Plemelj theorem evaluates the plasma dispersion function by equating the real-line integral over $v = (-\infty, \infty)$ with a closed contour integral in the complex plane. The residue theorem can then be applied, with the contour constructed such that no poles reside within it. This results in the common formula (written with $z = x + i\gamma$, to illustrate the importance of $\gamma$, or the imaginary part of $z$),
\begin{equation}
\label{eq:plemelj}
    \int_{-\infty}^\infty \frac{f(v)}{v- x \pm i \gamma}dv  =
    \mp i \pi f(x) +\mathcal{PV} \int_{-\infty}^{\infty} \frac{f(v)}{v - x} dv,
\end{equation}
which is only valid for the limit $\gamma \rightarrow 0$. The utility of the Plemelj formula in Eq. \ref{eq:plemelj} is the ease in computing the imaginary component as half of the residue of the pole. In contrast, evaluating the principal value integral, $\mathcal{PV}$, for the real part becomes more difficult than the original integral as the integration path is now directly through the pole (this is readily seen by noting which integrals in Eq. \ref{eq:plemelj} have a $\gamma$). An analytic solution of the principal value integral is available for the Maxwellian distribution function in terms of the imaginary error function and the kappa distributions in terms of hypergeometric functions. For any other distribution function, the principal value integral must be evaluated numerically.

Some numerical schemes for computing the principal value integral fold the integrand across the pole to create a one-sided Riemann sum with a special boundary term that handles the pole.\cite{nyiri2000}
The scheme works well if the evaluation of second derivatives near the pole is easy, but can fail to converge for numerically defined distribution functions. 
Another scheme involves numerically evaluating the principal value integral by integrating with a standard scheme up to a stand off distance from the pole;
then a correction term for the pole's residue is added based on a Taylor expansion of the integrand.\cite{palastro2010}
Both these\cite{nyiri2000, palastro2010} and other similar methods require well-behaved, smooth distribution functions. At their core, they also rely on the Plemelj formula in Eq. \ref{eq:plemelj}, which restricts the application to cases of poles that lie sufficiently close to the real line ($\im[z] \ll \re[z]$), and does not generalize for higher order poles.

This paper builds on the approach developed in Longley (2024)\cite{longley2024}, which foregoes the contour integral approach and keeps the integration along the real axis by choosing an adaptive velocity mesh with more refinement near the poles. This is justified by noticing that a pole such as $1/(v-x+i\gamma)$ can be rationalized if $\gamma \ne 0$, and the result can be separated into real and imaginary parts. For example, Eq. \ref{eq:gen-plasma-dispersion-function-one-pole} can be rewritten as two real valued integrals (with $z = x + i\gamma$):
\begin{align}
    \int_{-\infty}^\infty \frac{f(v)}{v- x \pm i \gamma}dv  &= \int_{-\infty}^\infty \frac{(v-x)f(v)}{(v- x)^2 + \gamma^2}dv \nonumber \\
        &\qquad\mp i\gamma\int_{-\infty}^\infty \frac{f(v)}{(v- x)^2 + \gamma^2}dv.
    \label{eq:will_wrong}
\end{align}

In this form, the denominators are strictly positive for all $v$ since $\gamma \ne 0$. Therefore, the generalized plasma dispersion function can be handled as an integral over the real line and does not inherently need an integration contour that closes in the complex plane. A numerical evaluation of Eq.~\ref{eq:will_wrong} can be accomplished by setting a velocity mesh that is more refined for $v \approx x$, with the Lorentzian shape of the integrand being used for automatic selection of the mesh. This approach avoided issues with standoff distance errors or derivatives near the pole and generalized well for higher order poles. However, it has convergence issues for either coarse velocity meshes or low collision rates that bring the pole near the real axis. The method described below keeps the integration on the real line, as in Eq.~\ref{eq:will_wrong}, and eliminates all of the disadvantages of each of these prior numerical schemes.

\section{Integration Scheme}  
\label{sec:method}
\subsection{Numerical Algorithm} 
\label{sec:method_subsection}

\begin{figure} 
    \centering
    \includegraphics[width=\linewidth]{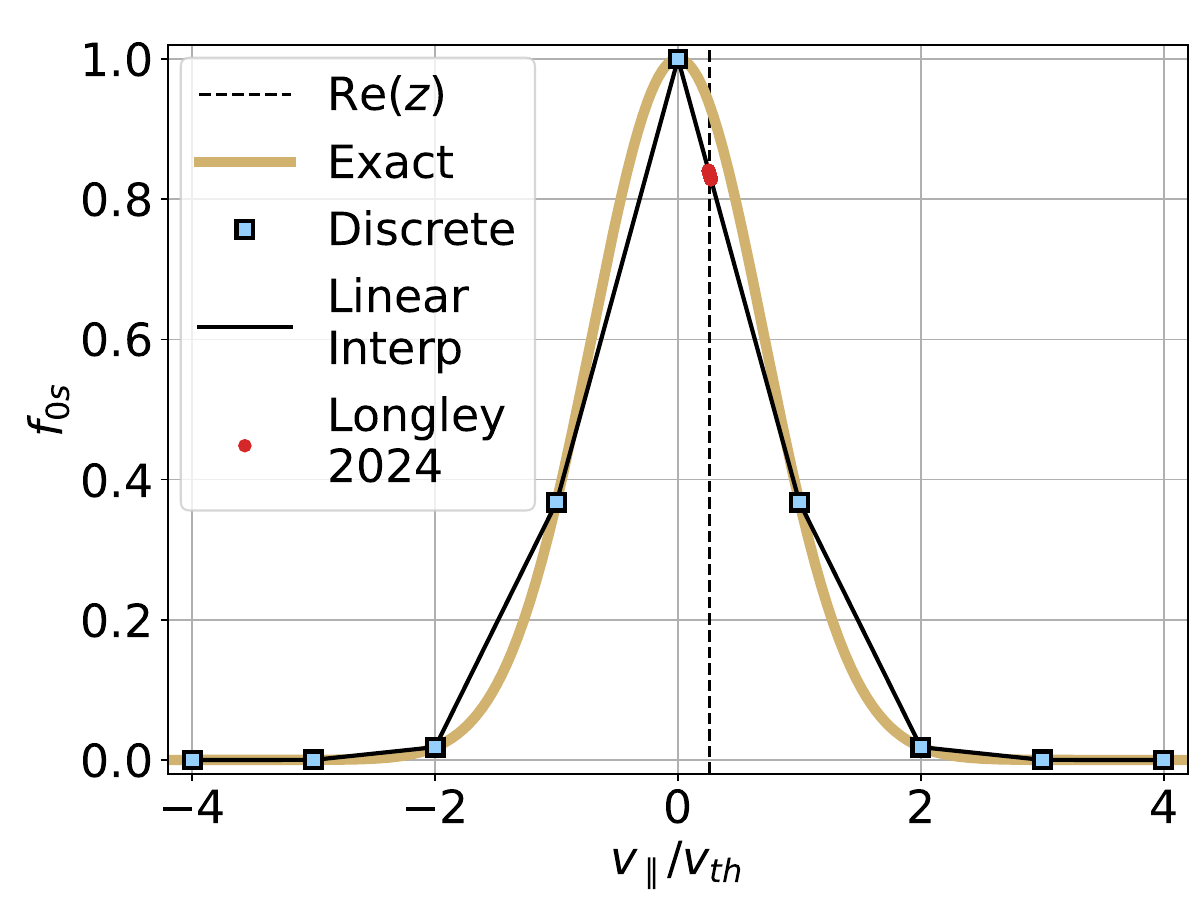}
    \caption{An example Maxwellian distribution (yellow line) represented by 9 ($m+1$) discrete points (black/blue squares) with a pole, $z$, whose real part is at $\pi/12$ (vertical dashed line).
    The black lines are the corresponding 8 ($m$) piecewise linear elements.
    The red dots correspond to the Longley (2024)\cite{longley2024} method used to refine the parallel velocity mesh about the pole.}
    \label{fig:discretization}
\end{figure}

To derive the numerical integration scheme, we consider first a distribution function in 1-D (Secs.~\ref{sec:thomson_app} and \ref{sec:thomson_results} generalize to 3-D, magnetized plasmas). 
Fig.~\ref{fig:discretization} shows an example distribution function (yellow curve), $f(v)$, that is discretized by $m+1$ equally spaced points (squares).
A normalized Maxwellian distribution is used for this example with spacing $\Delta v \equiv v_{j+1}- v_j = v_{th}$.
In our method, instead of considering a point-wise discretization, the distribution function can by discretized as a piecewise function, $\tilde{f}(v)$, with $m$ elements (black lines) as
\begin{equation}
	\tilde{f}(v) = 
	\begin{cases}
		\tilde{f}_0(v) & v_0 \leq v \leq v_1 \\
		\tilde{f}_1(v) & v_1 \leq v \leq v_2 \\
		\vdots  & \ \\
		\tilde{f}_j(v) & v_j \leq v \leq v_{j+1} \\
		\vdots & \\
		\tilde{f}_m(v) & v_m \leq v \leq v_{m+1} 
	\end{cases}.
	\label{eq:fLinearFull}
\end{equation}
The generalized plasma dispersion function can then be written as a summation of many integrals over the velocity mesh $v_j$. The infinite integration bounds are truncated so the integral is from $v_0$ to $v_{m+1}$, with those bounds chosen to capture the entire distribution function. Substituting the discretized distribution function, Eq.~\ref{eq:fLinearFull}, into the arbitrary pole generalized plasma dispersion function, Eq.~\ref{eq:gen-plasma-dispersion-function-sum-poles}, yields
\begin{equation}
Z(z_1, z_2, \ldots z_i, \ldots, z_N) = \sum_{j=0}^m \int_{v_{j}}^{v_{j+1}} \frac{\tilde{f}_j (v)}{\prod_{i=0}^N (v - z_i)^{r_i}}dv,
\label{eq:gen_split}
\end{equation}
which is the summation of the integral in each piecewise element.
Within each segment $[v_j, v_{j+1}]$, the distribution function is then linearly interpolated such that
\begin{equation}
	\tilde{f}_j (v) = a_j v + b_j,
	\label{eq:fLinear}
\end{equation}
where $a_j$ and $b_j$ are polynomial coefficients calculated using
\begin{align}
	a_j &= \frac{f(v_{j+1}) - f(v_{j})}{v_{j+1}-v_{j}}
	\label{eq:aCoeff} \\
	b_j &= f(v_{j}) - a_j v_{j},
	\label{eq:bCoeff}
\end{align}
where $f(v)$ here uses the known discrete values (square markers in Fig.~\ref{fig:discretization}).

The integral of the distribution function over arbitrary poles can then be approximated as the sum of integrals over the linear polynomials. 
Substituting Eq.~\ref{eq:fLinear} into Eq.~\ref{eq:gen_split} yields
\begin{equation}
Z(z_1, z_2, \ldots z_i, \ldots, z_N) = \sum_{j=0}^m \int_{v_{j}}^{v_{j+1}} \frac{a_j v + b_j}{\prod_{i=0}^N (v - z_i)^{r_i}}dv.
\label{eq:gen_split_linear}
\end{equation}


The utility in transforming the integral in this manner is that we no longer have an arbitrary distribution function. Instead, we have integrals of a linear function with poles. While the integral of an arbitrary distribution function does not have a general analytic solution, the integral of a linear function does.  Effectively, we exploit the fact that any numerical computation of any integral will eventually require a discretization of the integrand. We strategically choose to do the discretization at this step, as the resulting integral over $[v_j,v_{j+1}]$ has an analytic solution once the poles are specified. The numerical scheme is then completed by specifying the set of poles needed, then utilizing Mathematica or other symbolic math package to obtain analytic results for specific poles in Eq.~\ref{eq:gen_split_linear}.

As a set of examples, the numerical scheme is evaluated for pole integrals of:
\begin{align}
	&\int_{-\infty}^\infty \frac{f(v)}{v-z} dv
	\label{eq:p1} \\
	&\int_{-\infty}^\infty \frac{f(v)}{(v-z)^2} dv
	\label{eq:p2} \\
    &\int_{-\infty}^\infty \frac{f(v)}{(v-z)(v-z^*)} dv.
	\label{eq:pstar}  
\end{align}
These three integrals will be used for the calculation of the dielectric functions and Thomson scatter functions in Sections \ref{sec:thomson_app} - \ref{sec:thomson_results}.
The resulting discretized integrals of Eqs.~\ref{eq:p1} - \ref{eq:pstar} using Eq.~\ref{eq:gen_split_linear} are
\begin{widetext}
    \begin{equation}
        \int \frac{\tilde{f}(v)}{v - z} dv = 
    	\sum_{j=0}^m 
        \bigg[a_j \big(v-z\big) + \big[ a_j z + b_j ] \ln \big( v - z \big)\bigg]
        \Bigg|_{v_{j}}^{v_{j+1}}
    	\label{eq:p1j}
    \end{equation}
    \begin{equation}
    	\int \frac{\tilde{f}(v)}{(v - z)^2} dv = 
        \sum_{j=0}^m \bigg[
    	\frac{-a_j z - b_j}{v - z} + a_j \ln\big(v-z\big)
       \bigg] \Bigg|_{v_{j}}^{v_{j+1}}
    	\label{eq:p2j}
    \end{equation}
    \begin{equation}
    	\int \frac{\tilde{f}(v)}{(v - z)(v - z^*)} dv = 
         \sum_{j=0}^m \bigg[
    	- \frac{i \Big( \big[ a_j z + b_j \big] \ln \big[ v -z \big] - 
    		\big[ a_j z^* + b_j \big] \ln \big[ v -z^* \big] 
    		\Big)  }{2 \im(z)} \bigg] \Bigg|_{v_{j}}^{v_{j+1}}.
    	\label{eq:pstarj}
    \end{equation}
\end{widetext}

Eqs.~\ref{eq:p1j} - \ref{eq:pstarj} provide explicit expressions that can be implemented numerically. These expressions are similar in complexity to a trapezoid rule scheme, with the complex logarithm being the only added difficulty. In Python, the \verb*|numpy.log| function readily calculates the principal branch of the logarithm.

Eqs.~\ref{eq:p1j} - \ref{eq:pstarj} will satisfy most user's needs. For other sets of poles, the generalized plasma dispersion in Eq.~\ref{eq:gen-plasma-dispersion-function-sum-poles} can be evaluated with the same steps:
\begin{enumerate}
    \item Segment the integral over $(-\infty, \infty)$ into a summation of integrals over $[v_j, v_{j+1}]$ (obtaining Eq. \ref{eq:gen_split})
    \item Use Eqs.~\ref{eq:fLinear} - \ref{eq:bCoeff} to linearize the distribution function within each interval (obtaining Eq.~\ref{eq:gen_split_linear})
    \item Use a computer algebra package or integral table to obtain the analytic result for the integral of a linear function (e.g., Eqs.~\ref{eq:p1j} - \ref{eq:pstarj})
\end{enumerate}

In Sec.~\ref{sec:polynomial_method}, this scheme is further generalized to arbitrary order polynomial fits to the distribution function.

\subsection{Error Analysis} 
\label{sec:method_errors}
In this section, we show how the numerical accuracy of Eqs.~\ref{eq:p1j} - \ref{eq:pstarj} depends on the velocity resolution, $\Delta v$, and the collision/damping rate, $\gamma \equiv \im(z)$. We make this comparison using a  Maxwellian distribution function defined as
\begin{equation}
    f(v)= \exp(-\tilde{v}^2) \equiv 
    \exp \bigg( - \frac{v^2}{v_{th}^2} \bigg),
    \label{eq:gaussian}
\end{equation}
with normalized velocity $\tilde{v}$ and without the usual $\pi^{-1/2}$ scaling factor.
The exact solutions of Eqs.~\ref{eq:p1} - \ref{eq:pstar} using the distribution function from Eq.~\ref{eq:gaussian} are
\begin{align}
	&\int_{-\infty}^\infty \frac{\exp(\tilde{v}^2)}{\tilde{v}-z} d\tilde{v}
	\equiv P_1(z)  = \pi \exp(-z^2) \nonumber \\
 & \qquad \qquad \qquad \quad \times \Bigg[i \sgn \Big( \im \big[z \big] \Big)  - \erfi(z) \Bigg] 
	\label{eq:p1Exact} \\
 &\int_{-\infty}^\infty \frac{\exp(\tilde{v}^2)}{(\tilde{v}-z)^2} d\tilde{v}
	= - 2 \sqrt{\pi} - 2 P_1(z),
	\label{eq:p2Exact} \\
	&\int_{-\infty}^\infty \frac{\exp(\tilde{v}^2)}{(\tilde{v}-z)(\tilde{v}-z^*)} d\tilde{v}
	= \frac{i}{2\im(z)} \bigg[ P_1(z^*) - P_1(z) \bigg] ,
	\label{eq:pstarExact} 
\end{align}
respectively, where $P_1(z)$ is defined as the solution in Eq.~\ref{eq:p1Exact}.

\begin{figure*} 
    \centering
    \includegraphics[width=\linewidth]{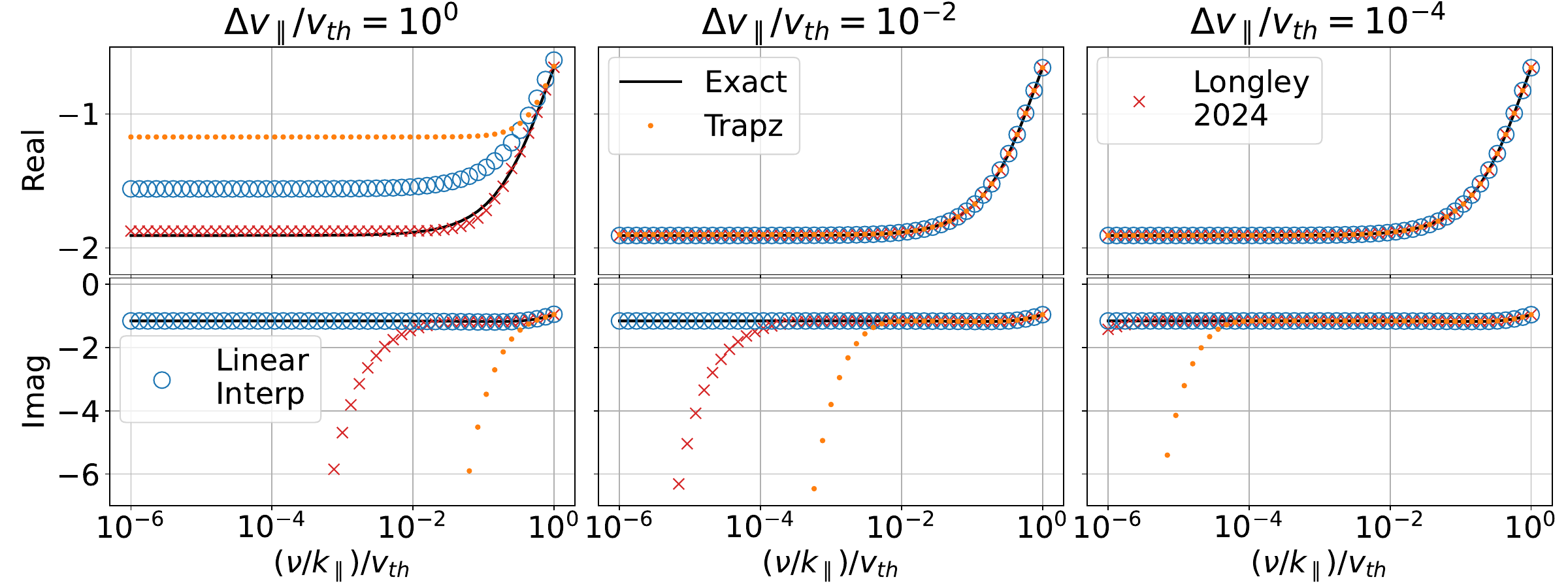}
    \caption{Plot of first order pole at $z$ (Eq.~\ref{eq:p1}) comparing the exact solution from Eq.~\ref{eq:p1Exact} (black line) to the linear interpolation integration method from Eq.~\ref{eq:p1j} (blue circles), the Longley (2024)\cite{longley2024} pole refinement method (red x's), and a trapezoid rule integration (orange dots).
    The $x$-axis represents the normalized collision rate and is the magnitude of the imaginary part of the pole $z$. 
    In other words, the pole is closer to the real axis on the left side of the plots (and therefore harder to integrate) and further from the real axis on the right side (and therefore easier to integrate). The top and bottom panels are the real and imaginary parts of the integrals, respectively.
    Each column of panels corresponds to a different resolution, $\Delta v$, of the distribution function (Eq.~\ref{eq:gaussian}).}
    \label{fig:p1Error}
\end{figure*}

\begin{figure*} 
    \centering
    \includegraphics[width=\linewidth]{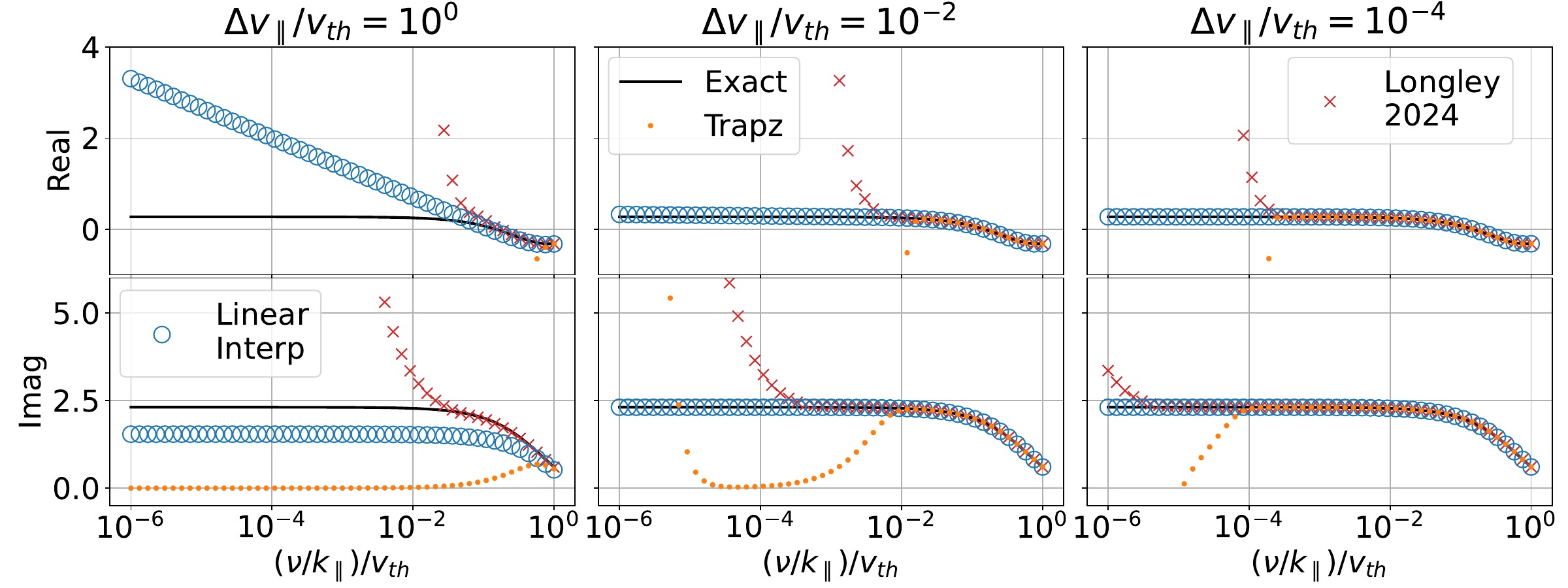}
    \caption{Same as Fig.~\ref{fig:p1Error} but for a second order pole at $z$ (Eq.~\ref{eq:p2}) using Eqs.~\ref{eq:p2j} and \ref{eq:p2Exact} for the linear interpolation and exact solution, respectively.}
    \label{fig:p2Error}
\end{figure*}

\begin{figure*} 
    \centering
    \includegraphics[width=\linewidth]{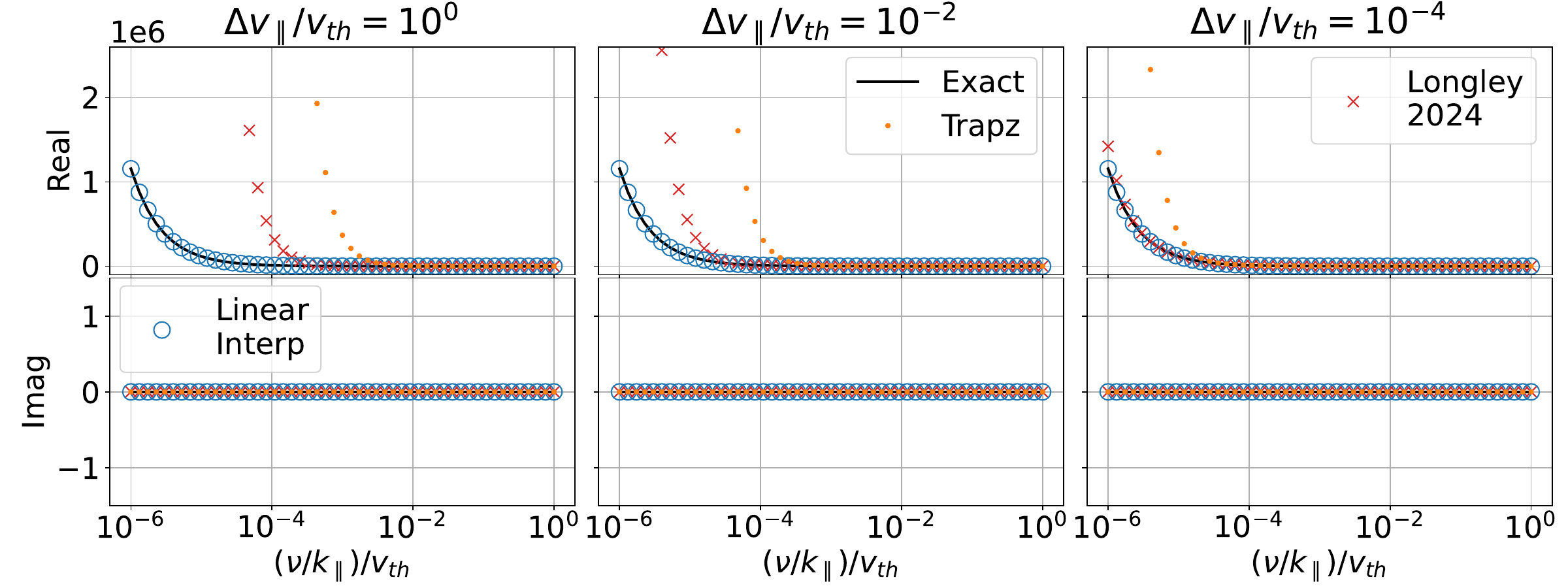}
    \caption{Same as Fig.~\ref{fig:p1Error} but for a set of first order poles at $z$ and $z^*$ (Eq.~\ref{eq:pstar}) using Eqs.~\ref{eq:pstarj} and \ref{eq:pstarExact} for the linear interpolation and exact solution, respectively.}
    \label{fig:pstarError}
\end{figure*}

Figs.~\ref{fig:p1Error} - \ref{fig:pstarError} compare the exact solutions, Eqs.~\ref{eq:p1Exact} - \ref{eq:pstarExact}, (black line) to the discretized solutions, Eqs.~\ref{eq:p1j} - \ref{eq:pstarj} (blue circles),
the Longley (2024)\cite{longley2024} pole refinement method (red x's), and a trapezoidal rule integration (orange dots).
For each figure, the Maxwellian distribution function is uniformly discretized with the columns showing velocity resolutions of $\Delta v/v_{th} = 10^0$, $10^{-2}$, and $10^{-4}$. The top and bottom panels show the real and imaginary parts of the solution, respectively.

For all calculations in this section, the location of the pole is set to $\re(z) = 1$ as this point has higher error than other tested values (units normalized by $v_{th}$). This choice is to show how useful the numerical scheme is for an unideal parameter regime. The imaginary part of the pole is set by plasma parameters, $\im(z) = (\nu_{coll}/k_\parallel)/v_{th}$, and is not always a free choice in calculating the generalized plasma dispersion function. To investigate the error associated with how close the pole is to the real axis, we continuously vary the normalized collision rate from $10^{-6}$ to $10^0$. Note that these parameter choices mean we calculate the case of $\im(z) = \re(z)$, which explicitly violates the Plemelj theorem's assumption of the pole being very close to the real line, $\im(z) \ll \re(z)$. The analytic solutions in Eqs.~\ref{eq:p1Exact} - \ref{eq:pstarExact} were derived without this assumption.

Fig.~\ref{fig:p1Error} shows the results for a first order pole at $z$ (Eq.~\ref{eq:p1}).
For the real part of the solution, the Longley (2024)\cite{longley2024} pole refinement method performs better than the linear interpolation method in the low resolution case ($10^0$), but neither method is accurate. At higher resolutions ($\Delta v = 10^{-2}, 10^{-4}$), all three integration methods converge to the exact solution.
For the imaginary part of the solution, the linear interpolation method converges at the coarsest resolution with the the other two methods improving at higher velocity resolution, but never converging for low collisionality. This shows that while the real part of the solution to the first order pole seems to be simple to solve with any method, the new linear interpolation method significantly outperforms the other two methods for the imaginary part of the solution.

Fig.~\ref{fig:p2Error} shows the results for the second order pole at $z$ (Eq.~\ref{eq:p2}).
Of all of the poles tested, the real part of the second order pole at $z$ is found to be the most difficult to capture, especially with the trapezoidal and Longley (2024)\cite{longley2024} methods.
The new linear interpolation method significantly outperforms the other two methods and fully calculates the solution starting at a resolution of $\Delta v = 10^{-2}$. 
At the highest resolution tested, trapezoidal integration and the Longley (2024)\cite{longley2024} pole refinement method diverge from the exact solution at a relatively large $(\nu/k_\parallel)/v_{th}$. 
For the imaginary part of the solution, the Longley (2024)\cite{longley2024} pole refinement method is the best method in the lowest resolution case, but still diverges from the analytic result. 
However, starting at higher resolutions, the linear interpolation method captures the exact solution whereas the other two methods diverge at lower collisionality.

Fig.~\ref{fig:pstarError} shows a set of first order poles at $z$ and $z^*$ (Eq.~\ref{eq:pstar}).
The conjugate poles can be multiplied to obtain a real valued integrand without a singularity. Therefore, the integral over the real line is necessarily real valued, as shown by all three numerical integration schemes in the bottom panels.
For the real part of the integral, the linear interpolation method captures the entire solution at the coarsest resolution, and the other two methods perform adequately at higher $\Delta v$ resolutions.

For all resolutions and pole types, all three integration methods converge to the exact solution as the pole moves further away from the real line. This is anticipated, as the imaginary part of the pole stabilizes the singularity in the integrand for real $v$, leading to a wide and easy to integrate function. Furthermore, each of the numerical integrations is more accurate as the velocity mesh is refined (smaller $\Delta v$). For the trapezoid rule, this dependence is based on a Riemann sum converging to an integral as $\Delta v \rightarrow 0$. For the other two schemes, the refinement does not effect the actual integration but instead sets how accurate a linear function can approximate the full distribution function (see Fig.~\ref{fig:discretization}). Therefore, the linear scheme developed in this paper is primarily limited by how well the distribution function is resolved in velocity space. 

\begin{figure}
    \centering
    \includegraphics[width=\linewidth]{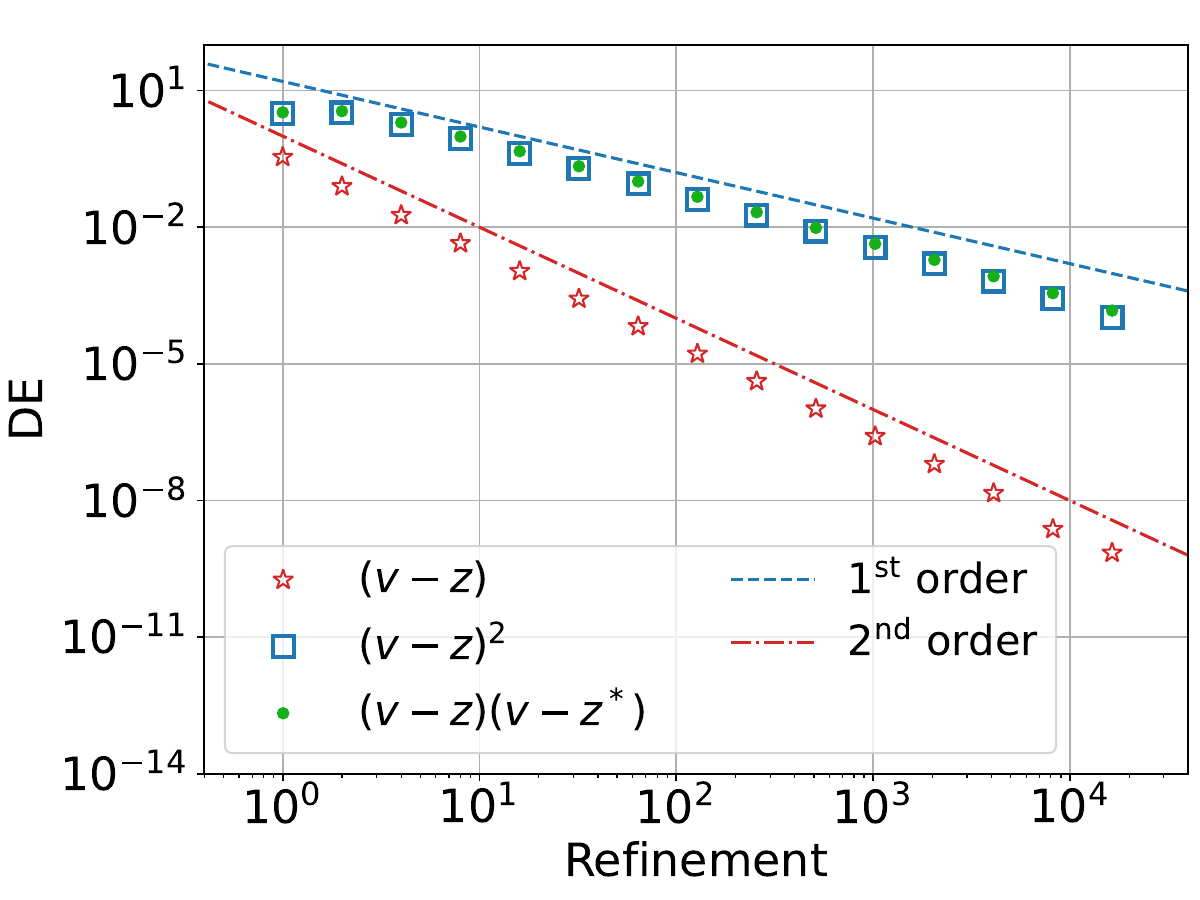}
    \caption{Discretization error versus velocity mesh refinement showing that the first order pole integration (Eq.~\ref{eq:p1j}) is second order accurate and the second order pole integrations (Eqs.~\ref{eq:p2j} - \ref{eq:pstarj}) are first order accurate.}
    \label{fig:order-accuracy}
\end{figure}

The relationship between resolution and accuracy is shown in Fig.~\ref{fig:order-accuracy}, which shows the discretization error versus the velocity mesh refinement for a Maxwellian distribution (Eq.~\ref{eq:gaussian}) with a pole at $z=1+10^{-6}i$.
The imaginary part of this pole is chosen for this calculation because it is the most difficult for all of the methods to capture, as shown in Figs.~\ref{fig:p1Error} - \ref{fig:pstarError}.
The discretization error is defined as the modulus of the difference between the discretized solution (Eqs.~\ref{eq:p1j} - \ref{eq:pstarj}) and the exact solution (Eqs.~\ref{eq:p1Exact} - \ref{eq:pstarExact}).
The coarsest and finest meshes have $\Delta v/v_{th} = 1$ and $\Delta v/v_{th}=2^{-14}$, respectively.
With a refinement factor of 2, the refinement varies from 1 to $2^{14}$.

The integration for the first order pole (Eq.~\ref{eq:p1j}) is second order accurate.
The integrations for the two second order poles (Eq.~\ref{eq:p2j} - \ref{eq:pstarj}) are first order accurate.
Future work may consider examining the relationship between choice of polynomial order for the piecewise functions in Eq.~\ref{eq:fLinearFull} (1 in our case) and the order of the pole.

In general, the new integration method outperforms both the trapezoid rule integration and the Longley (2024)\cite{longley2024} pole refinement method, with the trapezoidal integration being the worst method. The results from the linear interpolation scheme are less sensitive to how close the pole is to the real line, and therefore this scheme is more robust and adaptable to all plasma parameters. A general guideline based on Figs.~\ref{fig:p1Error}-\ref{fig:pstarError} is that the linear interpolation method is highly accurate as long as $\Delta v/v_{th} \le 10^{-2}$.

\subsection{Extension to Generalized Spectral Representations}
\label{sec:polynomial_method}

The linear interpolation method described in Sec. \ref{sec:method_subsection} can be generalized for any polynomial representation of the distribution function. 
Some discretization methods used for the Vlasov equation, such as 
finite volume\cite{minoshima2015,touati2021}, 
discontinuous Galerkin,\cite{shumlak2013,juno2018} 
and general spectral methods,\cite{delzanno2015,joglekar2020}
naturally use a polynomial basis.
In this section, we will show how a numerical algorithm to calculate the generalized plasma dispersion function can be derived for the discontinuous Galerkin (DG) method.

The DG method is chosen due to its generality.
Other discretization methods, such as the finite volume, Galerkin, or general spectral methods can be derived based on the foundation set for the DG method here.


Suppose the distribution function is discretized using the discontinuous Galerkin method with a $p$-th order polynomial representation in each cell:
\begin{equation}
	\tilde{f}_j(v)  =c_{0,j} + c_{1,j} v + c_{2,j} v^2 + \cdots +  c_{p,j} v^p
	=  \sum_{q=0}^p c_{q,j} v^q,
	\label{eq:arb_poly_dist} 
\end{equation}
where $c_{q,j}$ are the coefficents for the $q$-th order term within cell $j$.

Eq.~\ref{eq:arb_poly_dist} can be substituted into Eq.~\ref{eq:gen_split} and solved element-wise (i.e., over $j$).
As an example, the first order pole integral (Eq.~\ref{eq:p1}) would be
\begin{equation}
	\int \frac{\tilde{f}(v)}{v -z} dv = 
	\sum_{j=0}^m \sum_{q=0}^p c_{q,j}\int_{v_{j}}^{v_{j+1}} 
    \frac{v^q}{v -z} dv.
    \label{eq:gen-poly-example}
\end{equation}
A symbolic math package such as Mathematica is used to evaluate the integral.

Thus, for the set of poles from Eqs.~\ref{eq:p1} - \ref{eq:pstar}, the fully evaluated integrals using an arbitrary polynomial representation of the distribution function (Eq.~\ref{eq:arb_poly_dist}) are
\begin{widetext}
    \begin{equation}
    	\int \frac{\tilde{f}(v)}{v-z} dv = 
    	\sum_{j=0}^m \sum_{q=0}^p c_{q,j} \frac{v^{q+1}}{(q+1)z} \ _2F_1\bigg(1, q+1; q+2; \frac{v}{z}\bigg)
        \Bigg|_{v_{j}}^{v_{j+1}}
    	\label{eq:p1j_arb_poly}
    \end{equation}
    \begin{equation}
    	\int \frac{\tilde{f}(v)}{(v-z)^2} dv = 
    	\sum_{j=0}^m \sum_{q=0}^p c_{q,j} \frac{v^{q+1}}{(q+1)z^2} \ _2F_1\bigg(2, q+1; q+2; \frac{v}{z}\bigg)
        \Bigg|_{v_{j}}^{v_{j+1}}
    	\label{eq:p2j_arb_poly}
    \end{equation}
    \begin{equation}
    	\int \frac{\tilde{f}(v)}{(v-z)(v-z^*)} dv
    	= \sum_{j=0}^m \sum_{q=0}^p c_{q,j}
    	\frac{v^{q+1}}{2 i (q+1) |z|^2 \im(z)} 
    	\Bigg[z \ _2F_1\bigg(1, q+1; q+2; \frac{v}{z^*}  \bigg)  - z^* \ _2F_1\bigg(1, q+1; q+2; \frac{v}{z}  \bigg)\Bigg]
       \Bigg|_{v_{j}}^{v_{j+1}},
    	\label{eq:pstarj_arb_poly}
    \end{equation}
\end{widetext}
respectively,
where $_2F_1(a, b; c; z)$ is the hypergeometric function, which can be called using the \verb|scipy| package in Python with \verb|scipy.special.hyp2f1(a, b, c, z)|.

The integrated forms in Eqs.~\ref{eq:p1j_arb_poly} - \ref{eq:pstarj_arb_poly} can be simplified for other discretization methods. For example, a global Galerkin representation can be thought of as a discontinuous Galerkin method with only 1 cell. Therefore, the numerical scheme can be obtained by setting the end cell summation term as $m=0$. A finite volume representation can be thought of as a discontinuous Galerkin method with a polynomial order of 0. Therefore, the end polynomial summation term is $p=0$ in the above equations. This is also equivalent to setting $a_j=0$ and using the finite volume cell averaged value for $b_j$ from the linear interpolation method in Eq.~\ref{eq:fLinear}.

For spectral methods that use trigonometric or complex exponential bases, a similar analysis can be performed.
The resulting pole integrals are combinations of exponentials and the exponential integral: $\int_{-\infty}^{x}\frac{\exp(t)}{t}dt$
which can be called using the \verb|scipy| package in Python with \verb|scipy.special.expi(x)|.

\subsection{Advantages and Limits}

Compared to other methods for evaluating the generalized plasma dispersion function, the integration scheme developed in this paper has multiple distinct advantages.
\begin{enumerate}
    \item The integration scheme is comparable to a trapezoid rule integration of a non-singular function in terms of accuracy, speed, and ease of implementation
    \item The integration is not dependent on any frequency gridding, unlike Gordeyev integral calculations of Thomson scatter spectra.\cite{kudeki_milla2011}
    \item The integration is kept along the real line, and Eq.~\ref{eq:will_wrong} shows how the integral results in a complex value. This avoids conceptual difficulties with the Landau contour, analytic continuation, the Cauchy residue theorem, and the Plemelj theorem. 
    \item Numerically, there is no requirement that the velocity space must be evenly discretized. So long as sufficient initial resolution exists in regions of interest, this integration method will work with a non-uniform velocity mesh.
    \item By not using the Plemelj theorem, this scheme works equally well for high and low collisionality cases based on a comparison of the real and imaginary parts of the pole $z$. Additionally, the numerical challenge of principal value integrals is bypassed entirely.
    \item Velocity distributions obtained from experiments \cite{collinson2022, collinson2024} or simulations \cite{goodwin2018} are inherently discrete. The discretization and linear interpolation steps are therefore easy and natural to calculate when using the velocity resolution of the measured distribution.
    \item Accuracy can be improved by using polynomial or complex exponential fits to the distribution, shown in Section \ref{sec:polynomial_method}.
    \item While this paper only shows first or second order poles, the scheme is readily adapted to more complicated integrals. For example, the integration scheme works for the following 6$^\text{th}$-order pole integral:\cite{longley2024}
    \begin{equation}
        \int_{-\infty}^{\infty} dv \frac{f(v)}{(v-z_1)(v-z_1^*)(v-z_2)^2(v-z_2^*)^2}
    \end{equation}
    \item One strong advantage of the Plemelj approach to evaluating the integrals over a simple pole is that the imaginary part is succintly expressed as a residue of the distribution function at the pole (Eq. \ref{eq:plemelj}). For the case of low collisionality, $\im(z)~\ll~\re(z)$, the Plemlj form can be mixed with the numerical scheme providing the real part of the integral. For example, Eq.~\ref{eq:plemelj} can be rewritten as
\end{enumerate}
\begin{widetext}
\begin{align}
            \int_{-\infty}^\infty \frac{f(v)}{v- x \pm i \gamma}dv  &=
            \mp i \pi f(x) + \mathrm{Re} \left[ \int_{-\infty}^{\infty} \frac{f(v)}{v- x \pm i \gamma} dv \right]   
            \nonumber \\
        &= \mp i \pi f(x) + \mathrm{Re} \left[ \sum_{j=0}^m 
        \bigg[a_j \big(v-z\big) + \big[ a_j z + b_j ] \ln \big( v - z \big)\bigg]
        \Bigg|_{v_{j}}^{v_{j+1}} \right]
        \label{eq:plemelj_mixmatch}
    \end{align}
\end{widetext}

The primary limitation of the new integration scheme in Sec.~\ref{sec:method_subsection} is its dependence on the velocity resolution, $\Delta v$. This is a problem for all numerical integration methods. Figs.~\ref{fig:p1Error} - \ref{fig:order-accuracy} provide the guidance of $\Delta v < 10^{-2} v_{th}$ for Maxwellian distributions. However, for non-Maxwellian distributions, the concept of a thermal velocity may not apply, and there may also be bump-on-tail features that are narrow in velocity space. Therefore, it is not possible to provide a general guideline for the velocity resolution. Instead, this should be tested for each use case by evaluating the accuracy of $\int f(v) dv$ with a trapezoid rule or Riemann sum to find a suitable $\Delta v$ for convergence of the integral without any poles.

In addition, the velocity mesh extents must be such that they capture all of the distribution functions features. 
For the Maxwellian distributions used in Figs.~\ref{fig:p1Error} - \ref{fig:pstarError}, the velocity space extent is $\pm 4 v_{th}$, which fully captures the solution.


\section{Application to Thomson Scatter Theory}  
\label{sec:thomson_app}




The motivation for this study is to calculate the Thomson scattering spectra from non-Maxwellian plasmas. Thomson scattering is used with both ionospheric radars\cite{evans1969, beynon1978, Vallinkoski1988} and laboratory lasers\cite{gerry1966,froula,milder2019} to measure wave spectra of the plasma. These measured spectra are then fit to a forward model calculated using Maxwellian distribution functions, and an inverse process is used to find what plasma parameters create a best fit between the forward model and the measurement. This method is valuable for remote sensing the temperature and density of plasmas within small volumes. However, the fitting process is sensitive to the forward model used, and therefore being able to calculate the wave spectra for non-Maxwellian plasmas can extend the utility of this technique.  

In this section, we review the Thomson scattering theory from Froula and Sheffield (2011)\cite{froula} which will be solved in the following section for several non-Maxwellian distributions. In the Froula and Sheffield (2011)\cite{froula} framework (see also Refs.~\onlinecite{salpeter1960, salpeter1961, perkins1965}), the forward model for Thomson scatter experiments is obtained from solving the kinetic plasma equations. This results in explicit integrals over complex valued poles that will be numerically solved with the algorithm in Section \ref{sec:method_subsection}. We note that there are other approaches to solving for the Thomson scatter spectra using Gordeyev integrals.\cite{kudeki_milla2011, hagfors1961, farley1961} The Gordeyev integral frameworks all rely on a Kramers-Kronig relation to specify the real and imaginary parts of an integral in terms of each other, which is equivalent to applying the Plemelj theorem. The numerical integration in Section \ref{sec:method_subsection} is not compatible with these Gordeyev frameworks, as it effectively replaces the Kramers-Kronig relations step that is inherent to those derivations.

From Froula and Sheffield (2011)\cite{froula}, the spectral density, $S$, of Thomson scattered waves for arbitrary distribution functions is
\begin{equation}  
	S(\omega,\mathbf{k}) = 2 \Big| 1 - \frac{\chi_e}{\epsilon}\Big|^2 M_e
	+ 2\Big|\frac{\chi_e}{\epsilon}\Big|^2 \sum_i M_i,
	\label{eq:scattering-spectra}
\end{equation}
where $\omega$ is the Doppler shifted angular frequency and $\mathbf{k}$ is the wavenumber of the Bragg scattered incident wave. The modified distribution function, $M_s$, physically describes the scatter from species $s$ in the plasma if it were a free gas, with no collective effects. In Eq.~\ref{eq:scattering-spectra}, the summation term represents the summation of the modified distribution functions for all ion species present in the plasma. In a magnetized, collisional plasma, the modified distribution function is
\begin{equation}
	M_s = \frac{-\displaystyle\frac{|U_s|^2}{\nu_s} + \nu_s \displaystyle\sum_{n=-\infty}^\infty \displaystyle\int 
	\frac{J_n^2\Big( \dfrac{k_\perp \text{v}_\perp}{\Omega_{cs}} \Big) f_{0s}(\mathbf{v}) d\mathbf{v}^3}
	{(\omega - k_\parallel v_\parallel - n\Omega_{cs})^2 + \nu_s^2}}{|1+U_s|^2},
	\label{eq:Ms}
\end{equation}
where $\nu_s$ is the collision frequency and $v$ is the velocity. The subscripts $\perp$ and $\parallel$ are the vector components of $v$ or $k$, either perpendicular or parallel to the the magnetic field. The convention is used where $F_{0s} = n_{0s}f_{0s}$ is the zeroth order distribution function, with $\int dv^3 f_{0s}=1$. The gyrofrequency is $\Omega_{cs} = q_sB/m_s$ where $q_s$ is the charge, $B$ is the magnetic field strength, and $m_s$ is the particle mass. $J_n$ is the $n$-th order Bessel function of the first kind. $U_s$ is a collisional term defined as
\begin{equation}
	U_s = i \nu_s \sum_n \int 
	\frac{J_n^2\Big( \dfrac{k_\perp v_\perp}{\Omega_{cs}} \Big)}
	{\omega - k_\parallel v_\parallel - n\Omega_{cs} - i\nu_s}  
	f_{0s}(v)   d\mathbf{v}^3.
	\label{eq:Us}
\end{equation}

The collective behavior of the plasma is described by the susceptibilities of each species, $\chi_s$, and the longitudinal dielectric function, $\epsilon$, which is defined as
\begin{equation}
	\epsilon = 1 + \chi_e + \sum_i \chi_i.
	\label{eq:dielectric}
\end{equation}

The magnetized, collisional susceptibilities are calculated as
\begin{align}
	\chi_s &= \frac{\omega_{ps}^2}{k^2(1+U_s)} \sum_{n=-\infty}^\infty \int 
	\frac{J_n^2\Big( \frac{k_\perp v_\perp}{\Omega_{cs}} \Big)}
	{\omega - k_\parallel v_\parallel - n\Omega_{cs} - i\nu_s} \nonumber \\  
	&\qquad \qquad \times \Bigg(k_\parallel \frac{\partial f_{0s}}{\partial v_\parallel}
	+ \frac{n \Omega_{cs}}{v_\perp} \frac{\partial f_{0s}}{\partial v_\perp} \Bigg)  d\mathbf{v}^3 ,
	\label{eq:chis}
\end{align} 
where $\omega_{ps}=\sqrt{n_se^2/m_s\varepsilon_0}$ is the plasma frequency,
$e$ is the elementary electric charge, and $\epsilon_0$ is the permittivity of free space.

For most stable, magnetized plasmas, the distribution function will have an azimuthal symmetry about the magnetic field. This allows the integrals in Eqs.~\ref{eq:Ms}, \ref{eq:Us}, and \ref{eq:chis} to be evaluated in cylindrical coordinates, with $f_{0s}$ having no dependence on the azimuth angle $\phi$. The integral of an arbitrary azimuthally symmetric function, $g(\mathbf{v})$, over all velocity space is
\begin{equation}
	\int g(\mathbf{v}) d\mathbf{v} = 2\pi \int_0^\infty  dv_\perp v_\perp  \int_{-\infty}^\infty g(v_\perp, v_\parallel)  dv_\parallel,
	\label{eq:cylind_int_2pi}
\end{equation}

Finally, in simplifying Eqs.~\ref{eq:Ms}, \ref{eq:Us}, and \ref{eq:chis}, we use the shorthand $z$ where
\begin{equation}
	z = \frac{\omega - n\Omega_{cs}- i \nu_s}{k_\parallel},
	\label{eq:z}
\end{equation}
is the Landau or cyclotron resonant velocity.

Using the azimuthal symmetry (Eq.~\ref{eq:cylind_int_2pi}) and the resonant velocities (Eq.~\ref{eq:z}), Froula and Sheffield (2011)\cite{froula} shows that the modified distribution (Eq. \ref{eq:Ms}) and collisional terms (Eq. \ref{eq:Us}) simplify to 
\begin{align}
	M_s &= \frac{\nu_s}{|1+U_s|^2} 
	\Bigg[ -\frac{|U_s|^2}{\nu_s^2} 
	+ \sum_{n=-\infty}^\infty \int_0^\infty dv_\perp 
    v_\perp J_n^2 \bigg( \frac{k_\perp v_\perp}{\Omega_{cs}} \bigg) \nonumber \\
	&\qquad \qquad \times \frac{1}{k_\parallel^2} 
    \underbrace{\int_{-\infty}^\infty 
    \frac{f_{0s}(v_\perp, v_\parallel)}{(v_\parallel - z)(v_\parallel - z^*)} dv_\parallel}_{
    \text{Eq.~\ref{eq:pstar}}
    }
	\Bigg],
	\label{eq:Ms_final}
\end{align}  
\begin{align}
	U_s &= - \frac{2 \pi i \nu_s}{k_\parallel} \sum_{n=-\infty}^\infty 
	\int_0^\infty dv_\perp v_\perp J_n^2 \bigg( \frac{k_\perp v_\perp}{\Omega_{cs}} \bigg) \nonumber \\
	& \qquad \qquad \times \underbrace{\int_{-\infty}^\infty  
	\frac{f_{0s}(v_\perp, v_\parallel)}{v_\parallel - z} dv_\parallel}_{\text{Eq.~\ref{eq:p1}}}
	\label{eq:Us_final}
\end{align}
where $z^*$ is the complex conjugate of $z$.

The same steps can be applied to simplify the susceptibilities in Eq. \ref{eq:chis}. However, we anticipate future work to use measured or simulated distribution functions which are discrete, and not smooth. To avoid complications in calculating the derivatives of discrete distribution functions, we use integration by parts to move the derivative from the distribution function to the rest of the integrand. Working this out and using an identity for the derivative of Bessel functions, we obtain
\begin{widetext}
\begin{align}
	&\chi_s = 
	\frac{2\pi \omega_{ps}^2}{k^2 (1+U_s)}	\sum_n
	\Bigg\{ - \int_0^\infty dv_\perp v_\perp J_n^2 \bigg( \frac{k_\perp v_\perp}{\Omega_{cs}}  \bigg) 
	\underbrace{\int_{-\infty}^\infty \frac{f_{0s}(v_\perp, v_\parallel)}{(v_\parallel-z)^2} dv_\parallel}_{\text{Eq.~\ref{eq:p2}}}  \hfill \nonumber \\
	& \qquad \qquad \qquad \qquad + 
    \frac{nk_\perp}{k_\parallel} \int_0^\infty dv_\perp J_n \bigg( \frac{k_\perp v_\perp}{\Omega_{cs}}  \bigg)
	\Bigg[ 
	J_{n-1} \bigg( \frac{k_\perp v_\perp}{\Omega_{cs}}  \bigg)
	-J_{n+1} \bigg( \frac{k_\perp v_\perp}{\Omega_{cs}}  \bigg)
	\Bigg]
	\underbrace{\int_{-\infty}^\infty \frac{f_{0s}(v_\perp, v_\parallel)}{v_\parallel-z} dv_\parallel}_{\text{Eq.~\ref{eq:p1}}}
	\Bigg\}.
	\label{eq:chis_final}
\end{align}
\end{widetext}

The integrals in Eqs.~\ref{eq:Ms_final} - \ref{eq:chis_final} are a nesting of $v_\perp$ and $v_\parallel$ integrals, with the $\phi$ integral already evaluated to obtain the leading $2\pi$ factor. 
The $v_\parallel$ integrals correspond to the poles (Eqs.~\ref{eq:p1} - \ref{eq:pstar}) used in Sec.~\ref{sec:method_subsection}.

The underbraced terms in Eqs.~\ref{eq:Ms_final} - \ref{eq:chis_final} show how the numerical integration in Sec.~\ref{sec:method_subsection} is naturally extended to a 3-D plasma. First, the 3-D kinetic equations are solved in a desired coordinate system. This results in a nested volume integral, where there will be a pole along one velocity coordinate based on the product $\mathbf{k}\cdot \mathbf{v}$ and the coordinate system. The outer integrals (e.g., the $v_\perp$ integral in cylindrical coordinates) are discretized using a standard integration scheme, such as the trapezoid rule. Then for each discrete value of $v_\perp$, the $v_\parallel$ integral over the pole is performed (Eqs. ~\ref{eq:p1} - \ref{eq:pstar}) using the new integration scheme in Eqs.~\ref{eq:p1j} - \ref{eq:pstarj}. Note that when doing the linear interpolation of the distribution function in Eq.~\ref{eq:fLinear}, the coefficients $a_j$ and $b_j$ are calculated with $v_\perp$ being fixed. This interpolates the distribution function along the $v_\parallel$ direction only.

\section{Thomson Scattering Spectra Examples}
\label{sec:thomson_results}
In this section, the numerical integration scheme (Sec. \ref{sec:method}) is used to calculate the Thomson scattering spectra (Sec. \ref{sec:thomson_app}) for several types of distribution functions: 
Maxwellian,
kappa, 
super Gaussian, 
and toroidal 
distributions.

The physical parameters in the calculations for the Maxwellian, kappa, and super Gaussian distributions are chosen to be the same to more easily compare the differences.
The radar frequency is \SI{230}{\mega\hertz}, which corresponds to the EISCAT VHF and EISCAT 3D radars.\cite{wannberg2022} 
The magnetic field strength is \SI{2e-5}{\tesla} with an aspect angle chosen to be \SI{60}{\degree} from parallel to the magnetic field.
We use a single ion species of atomic oxygen, resulting in an ion mass of 16 amu.
Quasineutrality is assumed with ion and electron densities of \SI{e10}{\meter^{-3}}.
The ion and electron temperatures are \SI{1000}{\kelvin} and \SI{1200}{\kelvin}, respectively.
The ion and electron collision frequencies are \SI{10}{\hertz} and \SI{100}{\hertz}, respectively.

The choice of parameters for the toroidal distribution is different and chosen to match Ref.~\onlinecite{goodwin2018}.

The maximum summation term is chosen such that its individual effect on the final solution will be less than or equal to machine precision and thus the solution converges to the best possible numerical result.
This value changes depending on the species and type of distribution function.


\subsection{Maxwellian Distribution}

The accuracy and the validity of the numerical integration of Eqs.~\ref{eq:Ms_final} - \ref{eq:chis_final} is checked by comparing the results to the analytic solution with a Maxwellian distribution function. 
The Maxwellian distribution function in cylindrical coordinates is
\begin{equation}
    f_M(v_\perp, v_\parallel) = 
    \frac{\exp\bigg(-\frac{\big[v_\perp^2+v_\parallel^2\big]}{v_{th}^2}\bigg)}{\pi^{\frac{3}{2}} v_{th}^3}
    \label{eq:maxwellian}
\end{equation}
which defines $v_{th}=\sqrt{2KT_s/m_s}$. Froula and Sheffield (2011)\cite{froula} shows how this distribution function can be substituted into Eqs.~\ref{eq:Ms} - \ref{eq:chis} to obtain standard expressions. However, the derivation in Froula and Sheffield (2011)\cite{froula} simplifies the results by assuming the collisionality is low. As shown in Sec.~\ref{sec:method_errors}, the numerical integration works well even in the highly collisional regime. To test the numerical integration in the highly collisional regime, we therefore rederive the analytic solution for Eqs.~\ref{eq:Ms} - \ref{eq:chis} without any assumptions on the collision rate. The resulting expressions are 
\begin{widetext}
        \begin{equation}
        	U_s = 
        	\frac{i \nu_{s}}{k_\parallel v_{th,s}} 
        	\sum_n \exp(-k_\perp^2 \bar{\rho}_s^2) I_n (k_\perp^2 \bar{\rho}_s^2) 
        	\Big[ 2 i \sqrt{\pi} \exp(-y_n^2) - Z_M(y_n)\Big]
        	\label{eq:Us_exact}
        \end{equation}
        \begin{equation}
        	\chi_s = 
        	\frac{\alpha^2}{1+U_s} \frac{T_e}{T_s} 
        	\sum_n \exp(-k_\perp^2 \bar{\rho}_s^2) I_n (k_\perp^2 \bar{\rho}_s^2) 
        	\bigg[ 1 - \frac{\omega-i\nu_s}{k_\parallel v_{th,s}} 
        	\Big( 2 i \sqrt{\pi} \exp[-y_n^2] - Z_M[y_n] \Big)
        	\bigg]
        	\label{eq:chis_exact}
        \end{equation}
        \begin{equation}
        	M_s = - \frac{|U_s|^2}{\nu_s |1+U_s|^2} 
        	+ \frac{1}{k_\parallel v_{th,s} |1+U_s|^2}
        	\sum_n \exp \big(-k_\perp^2 \bar{\rho}^2_s \big)
        	I_n \big( k_\perp^2 \bar{\rho}^2_s \big)
        	\Bigg[ 2\sqrt{\pi} \re\bigg( \exp \Big[-y_n^2\Big] \bigg) - \im \bigg( Z_M\Big[y_n\Big] \bigg)  \Bigg],
        	\label{eq:Ms_exact}
        \end{equation}
\end{widetext}
where $I_n$ is the $n$-th order modified Bessel function of the first kind and $Z_m$ is the plasma dispersion function for a Maxwellian distribution, Eq.~\ref{eq:plasma-dispersion-function})\cite{fried1961} (also called the Dawson or Faddeeva function).
The remaining variables are defined as:
\begin{equation}
	y_n = \frac{\omega-n\Omega_{cs}-i\nu_s}{k_\parallel v_{th,s}}
\end{equation}
\begin{equation}
	\bar{\rho}_s = \frac{v_{th,s}}{\sqrt{2}\Omega_{cs}}
\end{equation}
\begin{equation}
	\alpha = \frac{1}{k \lambda_{D_e}},
\end{equation}
where $y_n$ is a normalized frequency, $\bar{\rho}_s$ is the average gyroradius, and $\alpha$ is the incoherent scattering parameter.

The collision rate, $\nu_s/k_\parallel v_{th,s}$, is important as it sets the imaginary part of $y_n$.
The low collisionality regime is recovered by substituting $y_n \rightarrow \re(y_n)$ (or equivalently $\nu_s \rightarrow 0$) in Eqs.~\ref{eq:Us_exact} - \ref{eq:Ms_exact}. 
This substitution does not change $U_s$ or $\chi_s$. In the modified distribution function, the term $\im(Z_M[y_n])$ goes to 0 in the low collisionality regime matching Ref.~\onlinecite{froula}.

\begin{figure} 
    \centering
    \includegraphics[width=\linewidth]{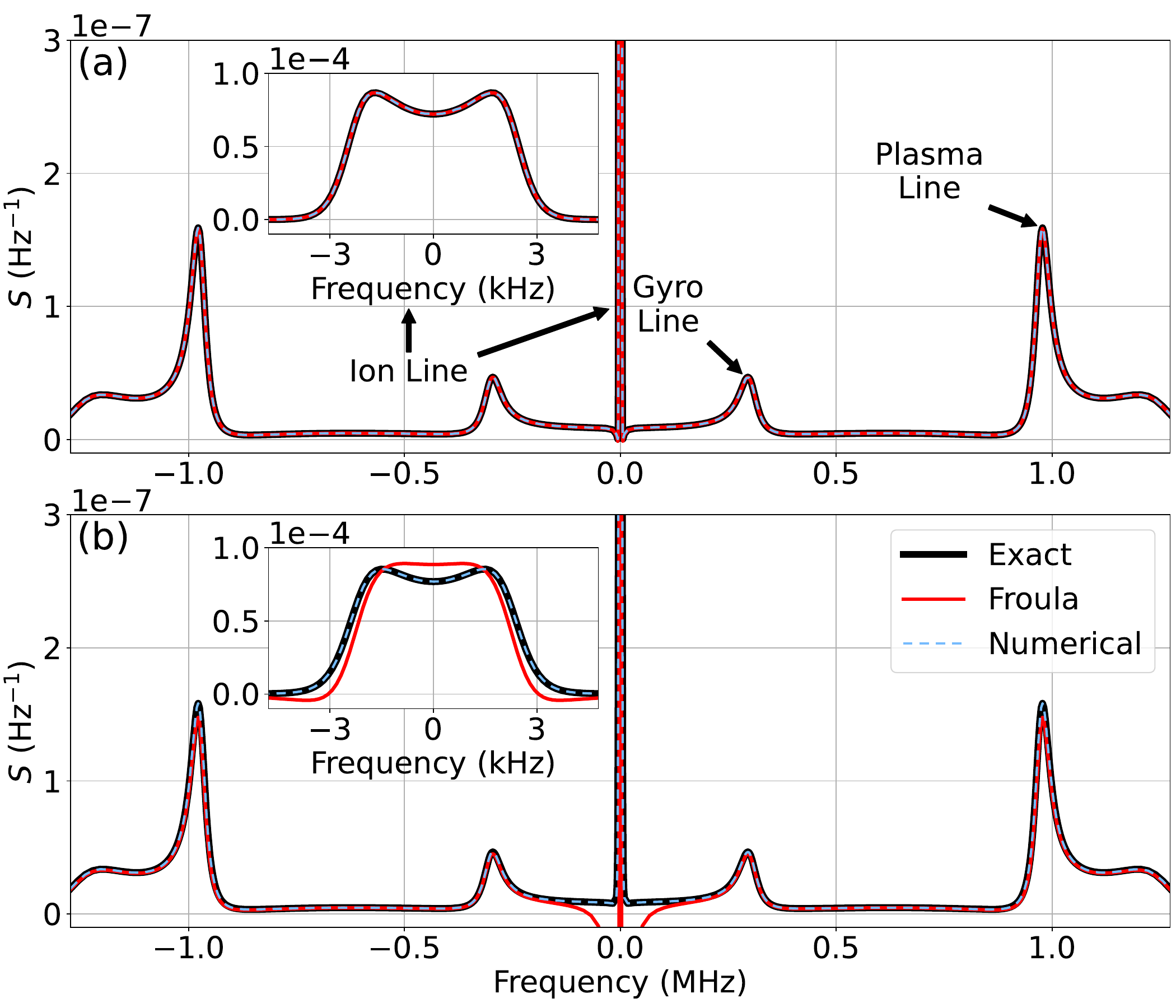}
    \caption{Thomson scatter spectra for Maxwellian distributions with low (a) and high (b) collisionality.
    For low collisionality, the numerical integration scheme (blue-dash) exactly reproduces the analytic solutions from Eqs.~\ref{eq:Us_exact} - \ref{eq:Ms_exact} (black line) and Froula and Sheffield (2011)\cite{froula} (red line).
    At high collisionality, the differences between the general analytic result in Eqs.~\ref{eq:Us_exact} - \ref{eq:Ms_exact} and the Froula and Sheffield (2011)\cite{froula} results are significant at low frequencies. 
    The numerical algorithm accurately reproduces the correct theory in the high collisionality regime.}
    \label{fig:maxwellian_spectrum}
\end{figure}

Fig.~\ref{fig:maxwellian_spectrum} shows the calculated Thomson scattering spectra for Maxwellian distributions.
Both the ion and electron contributions to the spectrum are calculated using the numerical integration technique from Sec.~\ref{sec:method}.
The maximum Bessel summation indices  
for the ions and electrons are 1000 and 17, respectively.
For both species, the velocity mesh has a resolution of $10^{-2}v_{th,s}$ in the perpendicular direction and $10^{-2.3}v_{th,s}$ in the parallel direction.
The velocity space extends from 0 to $4 v_{th,s}$ for the perpendicular direction and from $\pm 4 v_{th,s}$ for the parallel direction.

Panel (a) compares the numerical integration to the low collisionality solutions in Ref.~\onlinecite{froula}, and Panel (b) shows the results in the high collisionality regime. 
For low collisionality, the numerical integration (blue dashed line) accurately reproduces the standard analytic theory from Ref.~\onlinecite{froula} (red line) and the analytic results from Eqs.~\ref{eq:Us_exact} - \ref{eq:Ms_exact} (thick black line) across the entire frequency spectrum. 
For high collisionality, the numerical integration matches Eqs.~\ref{eq:Us_exact} - \ref{eq:Ms_exact} validating the numerical algorithm.
However, the standard theory from Froula and Sheffield (2011)\cite{froula} does not match either the numerical or newly derived analytical forms. 
The high frequency electron features are generally captured with a slight decrease of the plasma line peak.
At lower frequencies, however, there is significant deviation as shown in the inset plot of the ion line.
Furthermore, the scattering power becomes negative for a region above $\pm \SI{3}{\kilo \hertz}$ despite being a positive definite quantity.
Therefore, it is clear that for high collisionality regimes, the theory from Froula and Sheffield (2011)\cite{froula} is invalid with Eqs.~\ref{eq:Us_exact} - \ref{eq:Ms_exact} providing the correct solution without limitation on collisionality.

Panel (a) also provides annotations for the terminology of features often seen in ionospheric Thomson scattering.
The ion line is a double humped feature typically in the \si{\kilo\hertz} regime.
To match the differences in scales, an inset plot is shown that zooms in on the feature.
This is caused by scatter off of the ion acoustic mode.
At higher frequencies, the gyro line is seen and is caused by the electron gyration mode.
At even higher frequencies, the plasma line is seen and is caused by Langmuir modes.
This terminology will be used to discuss how these features change for different types of non-Maxwellian distributions.

\subsection{Kappa Distribution}

\begin{figure}
    \centering
    \includegraphics[width=\linewidth]{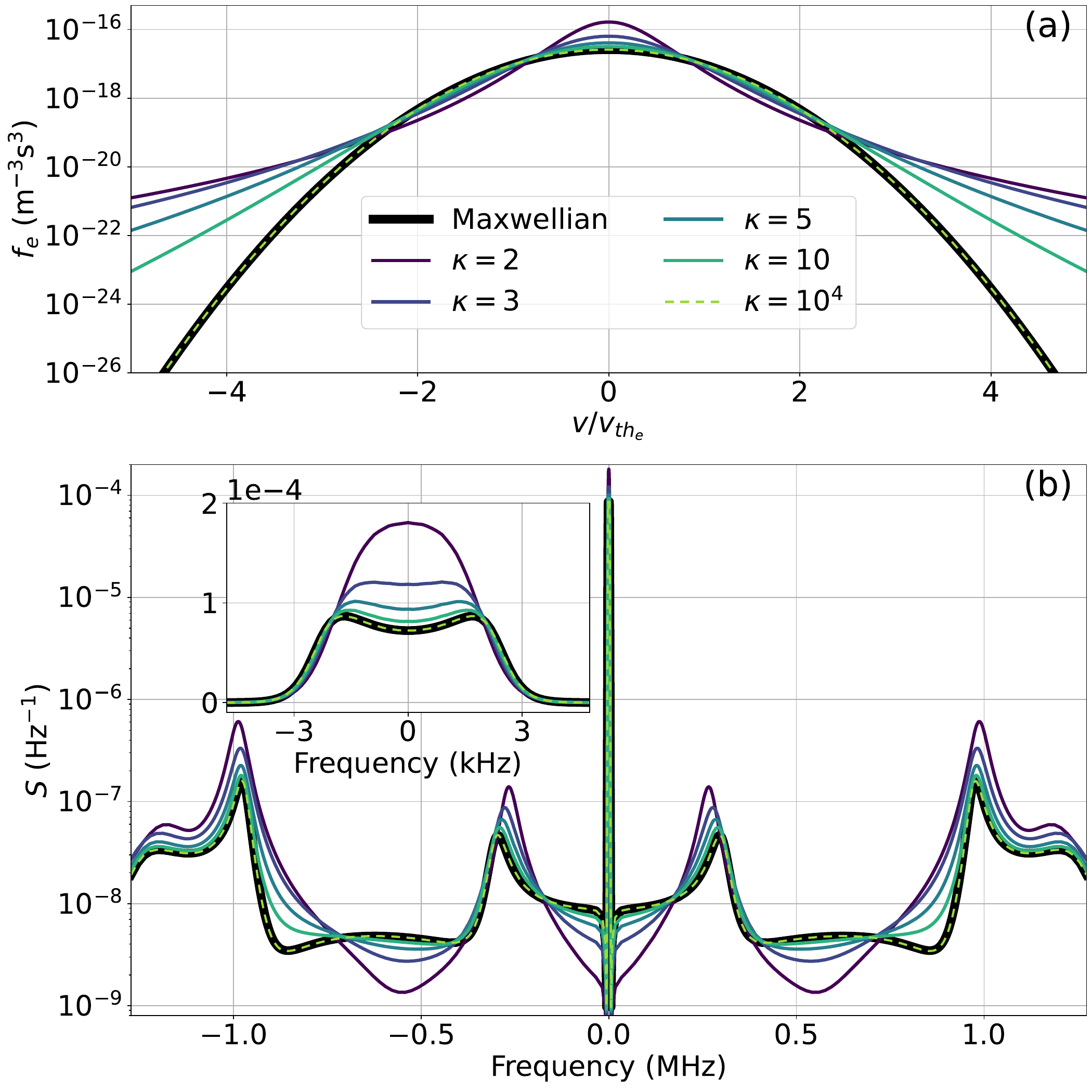}
    \caption{Panel (a) shows examples of kappa distributions (colored lines) and how they compare to a Maxwellian distribution (thick black line).
    Panel (b) shows the Thomson scattering spectra of the corresponding distribution functions.}
    \label{fig:kappa}
\end{figure}

Collisionless space plasmas such as the solar wind\cite{livadiotis2010} or within the magnetosheath,\cite{ogasawara2013} typically follow the kappa distribution.\cite{livadiotis2015}
The kappa distribution is characterized by a high energy tail population and is described analytically as\cite{livadiotis2015}
\begin{equation}
	f_{\kappa}(v_\perp,v_\parallel) = 
	\frac{\Gamma\big(\kappa+1\big)}{\Gamma\big(\kappa-\frac{3}{2}\big)}\frac{
	\Bigg( 1 + \frac{1}{\kappa-\frac{3}{2}} \frac{v_\perp^2+v_\parallel^2}{v_{th}^2}
	\Bigg)^{-\kappa-1}}{\pi^{\frac{3}{2}} v_{th}^3 \big(\kappa-\frac{3}{2}\big)^{\frac{5}{2}}},
	\label{eq:kappa_dist}
\end{equation}
where $\Gamma$ is the Gamma function 
and $\kappa$ is a parameter that must be $\geq 2$ and defines how much of a high energy tail population exists.
As $\kappa$ approaches $\infty$, the kappa distribution approaches the Maxwellian distribution (Eq.~\ref{eq:maxwellian}).
Fig.~\ref{fig:kappa}(a) shows a set of example kappa distributions (colored lines) for various $\kappa$ with a comparison to the equivalent Maxwellian distribution (thick black line).
The case of $\kappa=10^4$ is large and approximates the Maxwellian distribution.
For the lower $\kappa$ cases, a clear increase in the tail population is seen which will greatly affect the resulting Thomson scatter spectrum.

The Thomson scatter spectra are calculated assuming Maxwellian ions and kappa electrons described by the distributions shown in Fig.~\ref{fig:kappa}(a).
Therefore, the ion terms are calculated using the analytic theory for Maxwellians from Eqs.~\ref{eq:Us_exact} - \ref{eq:Ms_exact} and the electron terms are calculated numerically using the method from Sec.~\ref{sec:method}.
The maximum electron Bessel function summation index is 17.
The electron velocity mesh has a resolution of $10^{-2}v_{th,e}$ in the perpendicular direction and $10^{-2.5}v_{th,e}$ in the parallel direction.
The electron velocity space extends from 0 to $9 v_{th,e}$ for the perpendicular direction and from $\pm 9 v_{th,e}$ for the parallel direction.

The resulting spectra are shown in Fig.~\ref{fig:kappa}(b).
The spectrum for $\kappa=10^4$ matches that of the Maxwellian distribution as expected.

For low $\kappa$, the double peak structure of the ion line becomes one broad hump and shows an increases in scattered power.
This is caused by the kappa distribution increasing the electron population near 0 at low $\kappa$ causing a damping effect on the ion acoustic mode.

The high energy electrons in the low $\kappa$ distributions enhance the power in both the gyro and plasma lines.
Notably, the gyro line resonant frequency decreases while the plasma line resonant frequency increases. 
There are also regions with less scattered power in between all of these features.
The changes in the spectrum structure qualitatively agree with previous work.\cite{saito2000effects,enger2020model}

\subsection{Super Gaussian Distribution}

\begin{figure}
    \centering
    \includegraphics[width=\linewidth]{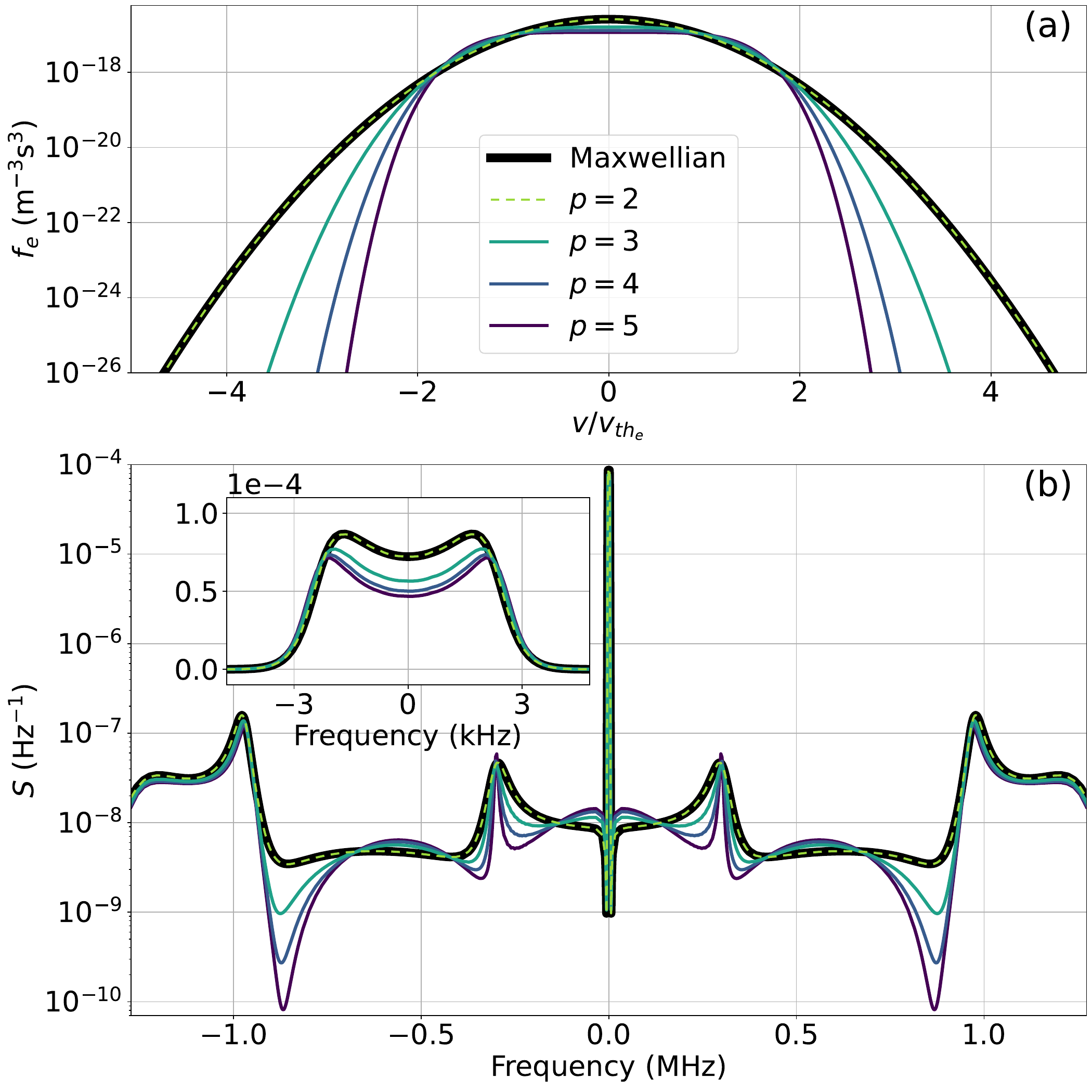}
    \caption{Panel (a) shows examples of super Gaussian distributions (colored lines) and how they compare to a Maxwellian distribution (thick black line).
    Panel (b) shows the Thomson scattering spectra of the corresponding distribution functions.}
    \label{fig:superGaussian}
\end{figure}

In high energy density plasmas in laser-driven inertial confinement fusion (ICF) experiments, electrons typically follow super Gaussian distributions.
These super Gaussian distributions are characterized by a lack of a tail population and an increase in particles near the thermal velocity.
In the case of ICF plasmas, this is because the inverse Bremsstrahlung heating from the lasers preferentially heats slow electrons to the thermal velocity.\cite{matte1988,milder2019}
The functional form of the super Gaussian distribution is
\begin{equation}
    f_{SG} = \frac{p}{4 \pi v_p^3 \Gamma(3/p)} 
    \exp \bigg[ - \frac{\big( v_\perp^2+v_\parallel^2 \big)^{\frac{p}{2}}}{v_p^p}   \bigg],
\end{equation}
where $p$ is a parameter that determines how small the tail population is 
and $v_p$ is the characteristic velocity defined as
\begin{equation}
    v_p = v_{th} \sqrt{ \frac{3 \Gamma (3/p)}{2 \Gamma (5/p)} }.
\end{equation}
Fig.~\ref{fig:superGaussian}(a) shows example plots of the super Gaussian distribution for various $p$.
When $p=2$, the super Gaussian distribution is the same as a Maxwellian distribution.
As $p$ increases, the tail population decreases and population near $v=0$ decreases slightly and becomes flatter;
this results in a slightly larger population at about 1 thermal velocity.

The Thomson scattering spectra are calculated assuming Maxwellian ions and super Gaussian electrons as described in Fig.~\ref{fig:superGaussian}(a).
Therefore, the ions are calculated using the analytic theory for Maxwellians from Eqs.~\ref{eq:Us_exact} - \ref{eq:Ms_exact} and the electrons are calculated using the numerical method from Sec.~\ref{sec:method}.
The maximum electron Bessel function summation index is 17.
The electron velocity mesh has a resolution of $10^{-2}v_{th,e}$ in the perpendicular direction and $10^{-2.3}v_{th,e}$ in the parallel direction.
The electron velocity space extends from 0 to $4 v_{th,e}$ for the perpendicular direction and from $\pm 4 v_{th,e}$ for the parallel direction.

The resulting spectra are shown in Fig.~\ref{fig:superGaussian}(b). 
The spectrum for $p=2$ matches that of the Maxwellian distribution, as expected.

As $p$ increases, the scattering power of the ion line decreases because of the decreased electron population near $v=0$.
In addition, because the electron distribution function is flatter near $v=0$, the two peaks of the ion line are shifted to higher frequencies with increasing $p$.

The gyro line resonant frequency does not appear to change with $p$;
the peak, however, becomes narrower for increasing $p$ and slightly increases in scattering power. The narrow peak is the result of fewer high energy electrons that typically damp the gyro line.
The plasma line is typically undamped, as there are few electrons with velocities high enough to resonate with the Langmuir mode. Therefore, the decrease in the suprathermal tail as $p$ increases leads to minimal changes in the plasma line.

\subsection{Toroidal Ion Distribution}

In the high latitude ionosphere, the effect of ion-neutral collisions, particularly the resonant charge exchange (RCE) reaction between O$^+$ and O, can significantly modify the ion distribution function.
In regions of high electric field, there is a drift between the ion and neutral populations.
A drifting neutral population that behaves in a Maxwellian manner will undergo some amount of RCE reactions.
This results in some of the drifting neutrals beginning to gyrating about Earth's geomagnetic field upon becoming ions resulting in an ion distribution in the shape of a torus.\cite{stmaurice1979,raman1981,winkler1992,goodwin2018}

An analytical approximation of such a toroidal distribution is\cite{stmaurice1979,raman1981,goodwin2018}
\begin{align}
	f_t &= \frac{
	\exp\bigg( - \frac{2 D^* v_\perp}{v_{th,\perp}} \bigg) I_0 \bigg(  \frac{2 D^* v_\perp}{v_{th,\perp}} \bigg) }{\pi^{3/2} v_{th,\parallel} v_{th,\perp}^{2}} \nonumber \\   
	& \qquad \qquad \times  \exp \bigg(  -\frac{v_\parallel^2}{v_{th,\parallel}^2} - \Big[ \frac{v_\perp}{v_{th,\perp} } - D^* \Big]^2  \bigg),
	\label{eq:toroidal_dist}
\end{align}
where $D^*$ is a distortion parameter related to the relative drift between the ions and neutrals, $I_0$ is the zeroth order modified Bessel function of the first kind, and $v_{th,\parallel}$ and $v_{th,\perp}$ are the parallel and perpendicular thermal velocities, respectively. 
When $D^*=0$ and $v_{th,\parallel}=v_{th,\perp}$, the toroidal distribution is the same as the Maxwellian distribution.
As $D^*$ increases, the torus becomes larger in velocity space.

The Thomson scattering spectra of a toroidal ion distribution and a Maxwellian electron distribution are calculated for varying aspect angles.
The Maxwellian electron terms are calculated using the analytic theory from Eqs.~\ref{eq:Us_exact} - \ref{eq:Ms_exact}.
The toroidal ion terms are calculated using the numerical method from Sec.~\ref{sec:method}.

The physical parameters are chosen to match those of Ref.~\onlinecite{goodwin2018} to make better comparisons.
The radar frequency is \SI{440}{\mega\hertz} based on the Advanced Moduler Incoherent Scatter Radars (AMISRs) located in Alaska and northern Canada.\cite{valentic2013}
The magnetic field is \SI{5e-5}{\tesla} with different spectra calculated for aspect angles of \SI{0}{\degree} (parallel to $\mathbf{B}$), \SI{20}{\degree}, \SI{30}{\degree}, \SI{60}{\degree}, and \SI{80}{\degree}. 
A single ion species of oxygen with mass 16 amu is used.
The ion and electron densities are quasineutral with a value of \SI{e11}{\meter^{-3}}.
The perpendicular and parallel ion temperatures are \SI{2000}{\kelvin} and \SI{1000}{\kelvin}, respectively.
The electron temperature is \SI{4000}{\kelvin}.
The ion and electron collision frequencies are \SI{1}{\hertz} and \SI{100}{\hertz}, respectively.
The distortion parameter, $D^*$, is 1.8.

The maximum ion Bessel function summation index is 2000.
The ion velocity mesh has a resolution of $10^{-2}v_{th,i}$ in the perpendicular direction and $10^{-2.3}v_{th,i}$ in the parallel direction.
The ion velocity space extends from 0 to $4 v_{th,i,\perp}$ for the perpendicular direction and from $\pm 4 v_{th,i,\parallel}$ for the parallel direction.

\begin{figure}
    \centering
    \includegraphics[width=\linewidth]{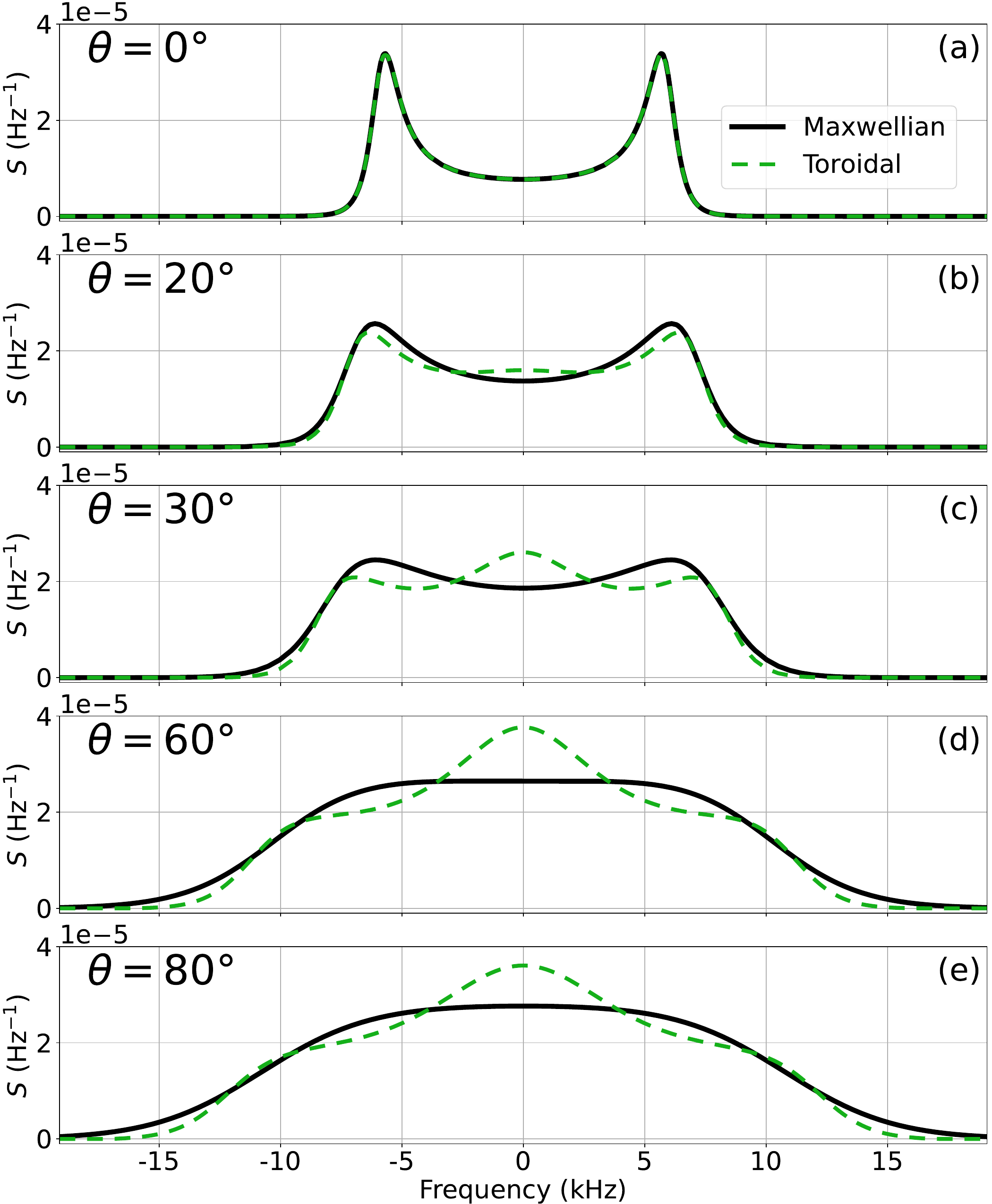}
    \caption{Thomson scattering ion line of a toroidal distribution (green dashed line) versus a Maxwellian distribution using the line of sight temperature (black line) for varying aspect angles. }
    \label{fig:toroidal}
\end{figure}

Fig.~\ref{fig:toroidal} shows the resulting spectra (green dashed lines) for varying aspect angles.
Since only the ions are non-Maxwellian, only the ion line is needed to examine the differences from a Maxwellian;
ions do not affect the high frequency electron gyration (gyro line) and Langmuir (plasma line) modes.

In addition, the black line represents the Maxwellian distribution based on the line of sight temperature.
For a normalized distribution function (i.e., zeroth moment of 1) with azimuthal symmetry and symmetry about the $v_\parallel=0$ plane, the temperature is defined by
\begin{equation}
    T_k = \frac{m M_{2_{kk}}}{k_B},
    \label{eq:los-temp}
\end{equation}
where $k$ represents the arbitrary line of sight direction, $m$ is the particle mass, $k_B$ is the Boltzmann constant, and $M_{2_{kk}} = \int v_k^2 f d \mathbf{v}$ is the second moment of the distribution function along the line of sight velocity direction $v_k$.

At $\theta=\SI{0}{\degree}$, which is parallel to the magnetic field, there is no difference between the toroidal and the Maxwellian spectra.
This is expected because the toroidal distribution comes about due to newly formed ions gyrating about the magnetic field.
This effect will not occur parallel to the field.

As the aspect angle increases to \SI{20}{\degree}, a small hump centered at 0 begins to form and the toroidal spectrum begins to deviate from the Maxwellian spectrum.
This matches previous results in the approximate angle at which the double humped spectrum transitions to a triple humped spectrum.\cite{goodwin2018}
At \SI{30}{\degree}, the middle hump continues to increase and a triple humped spectrum is seen.
At \SI{60}{\degree}, the middle hump continues to grow and reaches its peak value.
At \SI{80}{\degree}, the middle hump decreases slightly.

As the aspect angle increases, the line of sight Maxwellian spectra lose the double hump structure and becomes wider.
This is because the perpendicular temperature is greater than that of the parallel temperature.
Not only is this so because of how these two are initialized ($T_{i,\perp}=\SI{2000}{\kelvin} > T_{i,\parallel}=\SI{1000}{\kelvin}$) and used in Eq.~\ref{eq:toroidal_dist}, but also because the line of sight temperature shows an even stronger disparity between the temperatures.

Consider instead an anisotropic bi-Maxwellian distribution function 
\begin{equation}
    f_{bM}(v_\perp, v_\parallel) = 
    \frac{\exp\bigg(-\frac{v_\perp^2}{v_{th,\perp}^2}-\frac{v_\parallel^2}{v_{th,\parallel}^2}\bigg)}{\pi^{\frac{3}{2}} v_{th,\perp}^2 v_{th,\parallel}}.
    \label{eq:bimaxwellian}
\end{equation}
When taking the perpendicular line of sight temperature of a bi-Maxwellian, we would obtain the perpendicular temperature used to calculate $v_{th,\perp}$. 
However, the toroidal distribution is distorted by the $D^*$ parameter resulting in a wider distribution in the perpendicular direction;
in other words, the toroidal distribution has a higher perpendicular temperature in the perpendicular direction compared to the equivalent bi-Maxwellian distribution.
In our case, the line of sight ion temperature in the perpendicular direction, using Eq.~\ref{eq:los-temp}, is \SI{8254}{\kelvin} which is much larger \SI{2000}{\kelvin} which was used in Eq.~\ref{eq:toroidal_dist}.

\section{Summary and Conclusions}
\label{sec:conclusions}
In kinetic theory, we often have to take integrals of the distribution function over some set of complex poles.
For a Maxwellian distribution function, this typically is the plasma dispersion function.
For a non-Maxwellian distribution function, we instead have the generalized plasma dispersion function.
These are typically calculated using the Plemelj theorem with a principal value integral.
However, that method is only valid for low collisionality plasmas and only for one single order pole.
It is important to consider more poles (such as one second order pole or two first order poles as in this work) to properly calculate Thomson scattering spectra or examine nonlinear kinetic coupling.
To account for this, we define the generalized plasma dispersion function for arbitrary poles and develop a novel method for solving it.

An arbitrary distribution function is discretized into piecewise linear elements. 
The pole integrals in each cell can be calculated analytically and summed to obtain the total integral. This method is found to significantly improve computational time and accuracy at lower resolutions than previous numerical methods.
The method can also be scaled to work with piecewise arbitrary order polynomial elements; 
this allows for directly performing calculations using discontinuous Galerkin method expansion coefficient data.

This method is used to calculate the Thomson scattering spectra of non-Maxwellian distributions.
Typically, Thomson scattering inverse models assume Maxwellian distributions to obtain densities and temperatures.
However, when diagnosing a non-Maxwellian plasma, the Maxwellian assumption introduces signficant error.
Using our numerical method, we can calculate the forward model of an arbitrary distribution function that is azimuthally symmetric about the magnetic field.

We provide example spectra for kappa, super Gaussian, and toroidal distribution functions.
Future work can use this method to fully characterize a forward model for a particular distribution to improve Thomson scattering inverse models and spectral fitting routines.

\begin{acknowledgments}
This work was supported by NSF award AGS-2330254, and NJIT funds to L. V. Goodwin and C. R. Skolar.
\end{acknowledgments}

\section*{Data Availability Statement}
Readers may reproduce our results and also use the code for other applications. 
The general code and all spectra specific calculations are available online at the repository: \url{https://github.com/crskolar/ArbGenPlasmaDisp}.

\nocite{*}
\bibliography{reference}

\end{document}